\documentclass[12pt]{iopart}
\usepackage{verbatim}
\usepackage{soul}
\usepackage{xcolor}
\usepackage{lineno}

\usepackage{graphicx, epstopdf} 

\newcommand{\curl}{\nabla \times}
\newcommand{\dver}{\nabla \cdot}

\newcommand{\der}[3][]{ \frac{ \partial^{#1}{#3} }{ \partial{#2}^{#1} } }
\newcommand{\pder}[1]{\partial_{#1}}

\newcommand{\vect}[1]{\mathbf{#1}}

\begin{document}

\title{The c-equivalence principle and its implications for physics}
\author{Max Tran}
\vspace{10pt}
\begin{abstract}
	The c-equivalence principle, commonly accepted as true by most physicists, is the unstated assumption that $1/\sqrt{\epsilon_o \mu_o}$ equals the kinematic speed of light. Should someone prove the principle false, it would render the composition of two Lorentz transformations meaningless. It would also invalidate the second hypothesis of the Special Theory of Relativity in its strong form. This paper examine some other consequences for physics, should this principle be proven false and outline some experiments to determine light speed, which could falsify the principle and provide evidence for the ether.
\end{abstract}
\section{Introduction}
When James Clerk Maxwell formulated the equations of electromagnetism that now bears his name, he based them upon an ether/medium model characterized by three measurable quantities $\epsilon_o, \mu_o$, and $\sigma_o$.
Today we called these quantities respectively the electric permittivity, magnetic permeability, and electric conductivity of the vacuum.
Usually, the quantity $\sigma_o$ is set to zero and so terms having it as a factor disappear.
These $\sigma_o$-terms model the resistivity or dampening aspect of the medium to electromagnetic waves. Thus, in setting $\sigma_o$ to zero, we are modeling electromagnetic waves that are isolated from matter and can propagate without losing energy.
Without these damping terms, the speed of electromagnetic waves in Maxwell’s theory, denoted by $c_o$, is defined in terms of the other two quantities by $c_o = 1/\sqrt{\epsilon_o \mu_o}$.
The quantity $c_o$ is really the ratio between the electrostatic and electromagnetic units and is called Maxwell’s constant.
As to why we usually take it to be the velocity of electromagnetic waves, we are just continuing the practice of Maxwell. His words on the matter may provide some insight:
\begin{quote}
	It is manifest that the velocity of light and the ratio of the units are quantities of the same order of magnitude. Neither of them can be said to be determined as yet with such degree of accuracy as to enable us to assert that the one is greater than the other.
	It is to be hoped that, by further experiment, the relation between the magnitudes of the two quantities may be more accurately determined.
	In the mean time our theory, which asserts that these two quantities are equal, and assigns a physical reason for this equality, are not contradicted by the comparison of these results such as they are. \cite[p 436, Volume II]{maxwell}
\end{quote}

Yet according to the widely accepted Einsteinian special theory of relativity (STR), the medium ``does not exist’’ \cite{einstein1905} and so can not have properties like $\epsilon_o, \mu_o, \sigma_o$. Thus, STR can only define the speed of electromagnetic waves like light kinematically by $c$ = path length/duration of travel. Prof. Einstein and most researchers tacitly assumed that $c = c_o$, but there is no obvious reason or evidence that the two definitions should agree.\cite{heras, shadow} The assumption that $c = c_o$ is called the c-equivalence principle by Jose Heras, in the same vein as inertial mass is equivalent to gravitational mass.\cite{heras} Before Heras named it, Roberto Monti analyzed this assumption in his insightful article.\cite{monti} Even Herbert Ives, of the Ives-Stillwell experiment, did not believe that the kinematic speed of light has the same value in all inertial frames. He assumed length contraction and time dilation to get an elaborate formula for light speed as measured by moving detectors in his paper ``Light Signals on Moving Bodies as Measured by Transported Rods and clocks.’’ \cite{ives37B}

We will examine this principle and some consequences for physics should it be false, especially what it means to the Lorentz transformation. We will also outline some experiments that could invalidate the principle. In the first section, we look at the connection between the Lorentz transformation and the c-equivalence assumption. We then examine an ether model of light propagation and contrast it with STR and other theories in the third section. We briefly related three famous experiments used as evidence for the Lorentz transformation to the c-equivalent hypothesis in the fourth section. Lastly, we examine some experiments that could prove the c-equivalent proposition false.

\section{The Lorentz transformation and the c-equivalence principle}

Since the Lorentz transformation is the heart of the special theory of relativity, there are many derivations of it since the days of Lorentz and Einstein. In particular, Jean-Marc Levy-Leblond showed that the Lorentz transformation (LT) follows from five commonly accepted hypotheses.\cite{levy} These being (1) the principle of relativity by which is meant the covariancy of equations modeling physical phenomena, (2) the homogeneity of spacetime and linearity of inertial transformations, (3) the isotropy of space, (4) the requirement that the transformations form a mathematical group and (5) causality. Yet, in his derivation, the value of the velocity constant is unclear. Requiring the covariancy of Maxwell’s equations in the form derived by Heras will yield the value of this constant for electromagnetism. In vacuum, Maxwell’s equations, with a partial time derivative and not assuming the c-equivalence principle, have the forms:\cite{heras}
\begin{equation} \label{maxwell3}
\fl \qquad \dver \vect{E} = \frac{\rho}{\epsilon_o}, \quad \dver \vect{B} = 0, \quad
\curl \vect{E} = - \frac{c_o^2}{c^2} \der{t}{\vect{B}}, \quad \curl \vect{B} = \mu_0 \vect{J} + \frac{1}{c_o^2} \der{t}{\vect{E}}.
\end{equation}
Using the standard method to derive the wave equations from Maxwell’s equations on (\ref{maxwell3}) yields the same undampened wave equation in free space:
\begin{equation} \label{wave}
\fl \qquad \nabla^2 \vect{E} - \frac{1}{c^2} \der[2]{t}{\vect{E}} = 0, \quad \nabla^2 \vect{B} - \frac{1}{c^2} \der[2]{t}{\vect{B}} = 0,
\end{equation}
showing that $c$ is the speed of the EM wave.
The Lorentz transformation with the undetermined velocity parameter $U$ in the standard situation of two inertial frames moving with relative speed $v$ along their common x-axis is given by:
\begin{equation} \label{ltF1}
\fl \qquad x’ = \gamma (x - v t), \quad y’ = y, \quad z’ = z, \quad t’ = \gamma ( t - v x/U^2), \quad \gamma = (1 - v^2/U^2)^{-1/2}.
\end{equation}
By using the inverse Lorentz transformation and standard relations among differentials we get the following relations among the derivative operators:
\begin{equation} \label{ldiff}
\pder{x’} = \gamma \left( \pder{x} + \frac{v}{U^2} \pder{t} \right), \quad \pder{y’} = \pder{y}, \quad \pder{z’} = \pder{z}, \quad \pder{t’} = \gamma \left( \pder{t} + v \pder{x} \right).
\end{equation}
The most important equations to get the value of the LT velocity parameter are (\ref{maxwell3}) when the sources are zero. In fact, only one pair of equations from those equations involving the partials derivative with respect to $x$ and $x’$ are needed:
\begin{equation}\label{max0}
\qquad \pder{x}E_{y} - \pder{y}E_{x} = - \frac{c_o^2}{c^2} \pder{t}B_{z} \quad
\pder{z}B_{x} - \pder{x}B_{z} = \frac{1}{c_o^2} \pder{t}E_{y}.
\end{equation}
The corresponding equations for the moving frame are:
\begin{equation}\label{max1}
\qquad \pder{x’}E’_{y’} - \pder{y’}E’_{x’} = - \frac{c_o^2}{c^2} \pder{t’}B’_{z’} \quad
\pder{z’}B’_{x’} - \pder{x’}B’_{z’} = \frac{1}{c_o^2} \pder{t’}E’_{y’}.
\end{equation}
Using (\ref{ldiff}) in (\ref{max1}) and rearranging the equations yield:
\begin{equation}
\pder{x} ( \gamma E’_{y’} + \gamma \frac{v c_o^2}{c^2} B’_{z’} ) - \pder{y} E’_{x’} = - \frac{c_o^2}{c^2}
\pder{t} ( \gamma \frac{v c^2}{U^2 c_o^2} E’_{y’} + \gamma B’_{z’}), \label{max3}
\end{equation}
and
\begin{equation} \label{max4}
\pder{z}B’_{x’} - \pder{x} ( \gamma \frac{v}{c_o^2} E’_{y’} + \gamma B’_{z’} ) = \frac{1}{c_o^2} \pder{t}( \gamma E’_{y’} + \gamma \frac{v c_o^2}{U^2} B’_{z’} ) .
\end{equation}
The covariancy requirement of the equations forces (\ref{max3}) and (\ref{max4}) to have the same forms as (\ref{max0}). This makes the quantity within the first parentheses on the left-hand side of (\ref{max3}) equal to the one within the parentheses on the right-hand side of (\ref{max4}). Consequently, the coefficients of the $B’_{z’}$ terms must be equal:
\begin{equation}
\frac{\gamma v c_o^2}{c^2} = \frac{\gamma v c_o^2}{U^2},
\end{equation}
so that $U = c$. This process, if carried out in full using all the equations, would yield the transformations for the components of the $\vect{E}$ and $\vect{B}$ fields. Of course, if the c-equivalence principle is accepted, everything agrees.

One reason not to accept the c-equivalence principle is revealed in the analysis of time-of-flight experiments to determine light speed, which we examine next.

\section{Measuring Light Speed}

To provide context for what follows, we summarize the various theories of light before the wide acceptance of the special theory of relativity.
In the 19th century, the leading theory of light is that it is a wave requiring an all-pervading medium, the so-called ``luminiferous ether.’’
There were two main versions of the ether theory: (1A) the Earth and all other bodies moved through the ether with little interaction and (1B) the entrained ether, in which the Earth and other material bodies dragged the ether in their motions.
Movement through the ether as in (1A) would generate winds which would affect the speed of light depending upon the direction of the wind/movement.
Variant (1B) encompasses a spectrum, from partial to total entrainment. The totally entrained ether theories are often called local ether theories.
In a local ether theory, there is no relative motion of the Earth and the ether near the Earth’s surface where we live and conduct most of our experiments, hence no ether wind to detect.
A few scientists supported an alternative to the wave theory of light, namely the particle/ballistic theory, which requires no medium.

\begin{figure}
	
	\begin{center}
		{\includegraphics[scale=1.6]{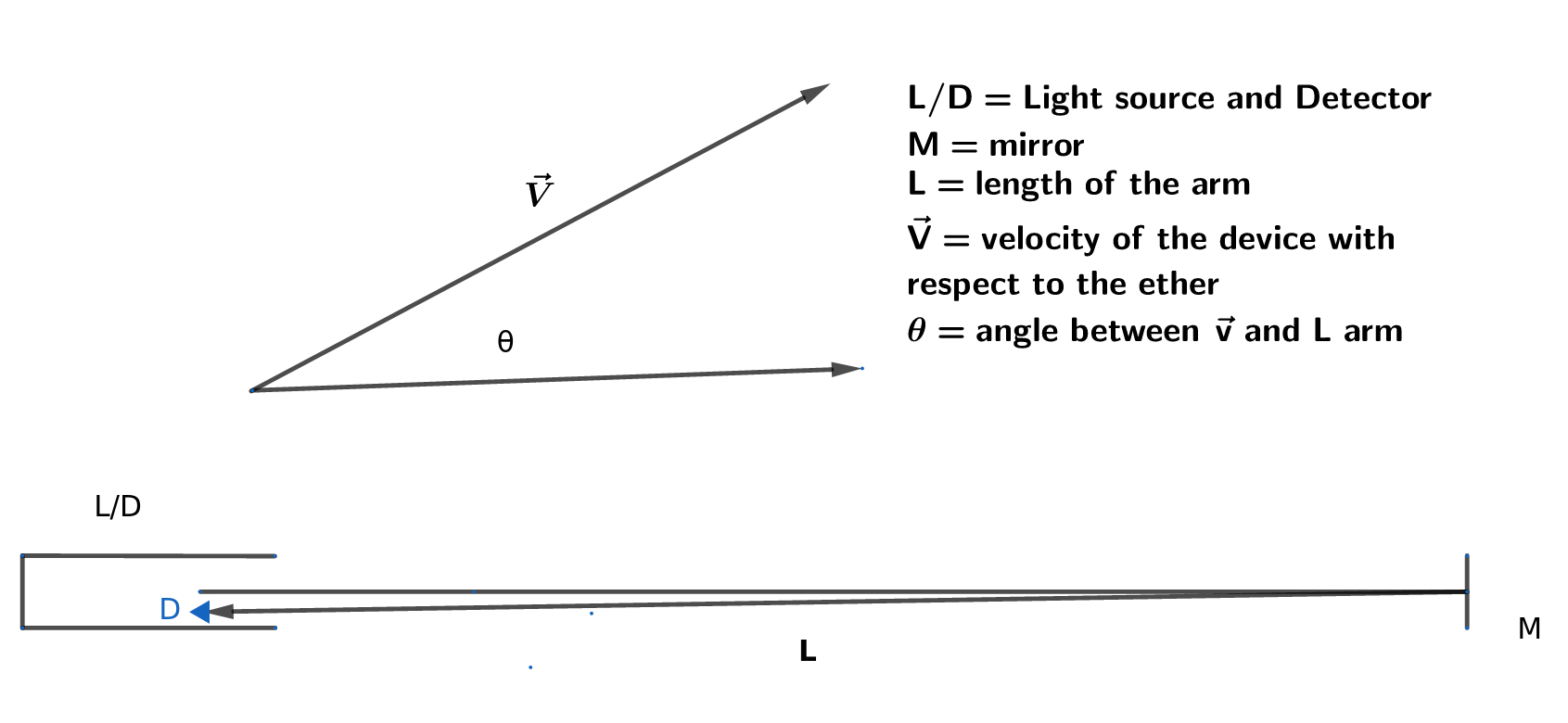}}
		\caption{A schematic drawing of a setup to measure the speed of light using time-of-flight methods. In actual devices, the light path may be more complicated. } \label{fig:Intfer}
	\end{center}
	
\end{figure} 

Assuming the existence of the ether, we now give a brief derivation of an equation relating the two speed parameters for an ideal setup to measure the speed of light by the time-of-flight method.
Let $S$ be a reference frame in which the electromagnetic speed of light $c_o$ equals the kinematic speed of light $c$. This frame is at rest relative to the ether.
In essence, the ether is defined to be the frame where $c= c_o = 1 / \sqrt{\epsilon_o \mu_o}$. Let $S_m$ be a frame moving with a constant nonzero velocity $\vect{v}$ relative to $S$. At rest in $S_m$ is a transceiver and reflector separated by a distance $L$.
In actual experiments, this frame has historically been the lab frame near the Earth’s surface.
We assumed the quantity $c_o$ to be invariant and unaffected by motion since it is a property of the ether. This assumption is akin to the constancy of light speed hypothesis in Einstein’s version of the special theory of relativity.

\begin{figure}	
	\begin{center}
		{\includegraphics[scale=1.6]{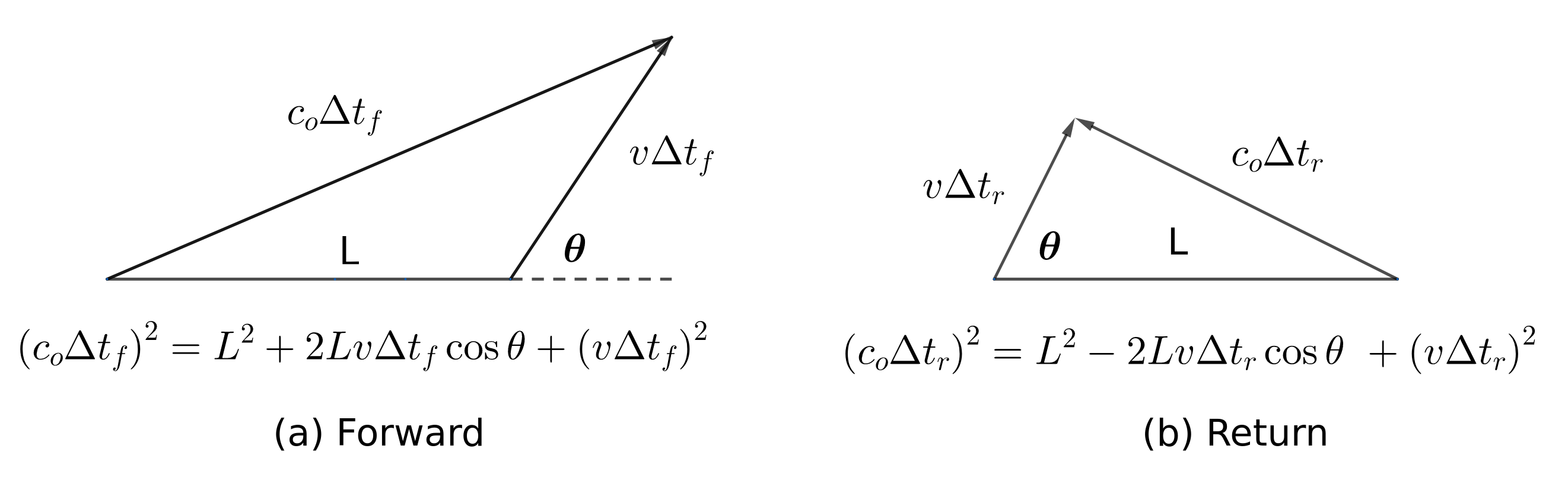}}
		\caption{Calculation of the optical paths for the forward and back directions of the light ray along the distance $L$ using the standard vector addition law. For $\theta = 0, \pi$, many would consider this situation as the addition of two velocities, but it is really just the calculation of the actually travel distance in the medium frame.} \label{fig:speedAdd}
	\end{center}	
\end{figure}

We need to find $c$ in the moving frame as the quotient of distance traveled and the duration of travel.
Referring to figure \ref{fig:speedAdd}, by solving the equation for the duration of travel in the forward direction of $L$ we get the forward travel duration to be
\begin{equation} \label{forward}
\Delta t_{f} = \frac{ L }{c_o(1 - \beta_o ^2)} [(1 - \beta_o^2 \sin^2 \theta)^{1/2} + \beta_o \cos \theta], \quad \beta_o = v/c_o,
\end{equation}
and for the return trip, the duration is
\begin{equation} \label{back}
\Delta t_{r} = \frac{L}{c_o(1 - \beta_o ^2)} [(1 - \beta_o^2 \sin^2 \theta)^{1/2} - \beta_o \cos \theta].
\end{equation}
Note that $\Delta t_{f} > \Delta t_{r}$, implying that the one-way speed of light relative to the moving detector and its rest frame is different in the forward and return paths even though the speed of light has the same value with respect to the medium in either direction.
We can understand the difference in the two durations as follows. In the forward path, the reflector is pulling away from the approaching light ray. Thus increasing the separation distance the ray must travel in the medium’s frame.
While on the return trip, the detector is approaching the ray while the ray is approaching it. This action decreases the separation distance between the ray and the detector in the medium’s frame.
If we superimpose the forward and return rays, fringe shifts due to any phase difference can be detected. This phase difference between the two rays is often called the generalized Sagnac effect and was confirmed in the experiments of Wang and collaborators.\cite{wangExp, wangExp2}
But many researchers consider measuring one-way speed light to be impossible since synchronization of clocks would be required.

The two-way speed of light is easier to do since it only requires one clock at the transceiver and no synchronization. Toward the expression for the two-way speed, the total duration of the back-and-forth travel along $L$ is
\begin{equation} \label{forback}
\Delta t_{fr} = \Delta t_{f} + \Delta t_{r} = \frac{ 2 L}{ c_o(1 - \beta_o ^2) } [(1 - \beta_o^2 \sin^2 \theta)^{1/2}].
\end{equation}
Thus, the two-way kinematic speed of light along $L$ according to the moving transceiver is
\begin{equation} \label{twowayspeed}
c = \frac{2 L}{\Delta t_{fr} } = \frac{ c_o ( 1 -\beta_o^2)}{( 1 -\beta_o^2 \sin^2 \theta)^{1/2} }.
\end{equation}
For frames moving with respect to the ether frame, \eref{twowayspeed} implies that $c_o > c$, falsifying the c-equivalence principle. Consequently, \eref{maxwell3} and \eref{wave} are the equations of the EM fields and waves in these frames.

If we accept the c-equivalence principle as true and the above model approximates reality, then (\ref{twowayspeed}) forces $v=0$, contradicting our working assumption of nonzero $v$.
Many people today accept the c-equivalence principle and reject the above classical Newtonian framework used in the analysis. Yet, even if length contraction and time dilation are assumed, the contradiction does not disappear as a quick calculation will reveal.
The two remaining possibilities are the entrained ether model that the velocity of the measuring device is zero relative to the local ether near the Earth’s surface, or that light travels like particles and is unaffected by the ether.

Unknown to most, there is a form of the entrained ether hidden within Einstein’s general relativity theory, as the following outline of his works will reveal.
In 1905, Albert Einstein put forth the hypothesis that the speed of light is constant in his special theory of relativity to derive the Lorentz transformation.
Once Professor Einstein developed the general theory of relativity, he made some interesting statements concerning the domain of validity of the constancy of light speed, a selection of which follows and can be found in his collected papers \cite{einsteincol}:
\begin{quote}
	{\em 1911:} If $c_o$ denotes the velocity of light at the coordinate origin, then the velocity of light $c$ at a point with a gravitation potential $\phi$ will be given by the relation $c = c_o (1 + \phi /c^2)$. The principle of the constancy of the velocity of light does not hold in this theory in the formulation in which it is normally used as the basis of the ordinary theory of relativity. [V3, D23, p 385]
	
	{\em 1913:} I have shown in previous papers that the equivalence hypotheses leads to the consequence that in a static gravitational field the velocity of light $c$ depends on the gravitational potential. This led me to the view that the special theory of relativity provides only an approximation to reality; it should apply only in the limit case where differences in the gravitational potential in the space-time region under consideration are not too great. [V4, D13, p 153]
	
	{\em 1916:} In the second place our result shows that, according to the general theory of relativity, the law of the constancy of the velocity of light in vacuo, which constitutes one of the two fundamental assumptions in the special theory of relativity and to which we have already frequently referred, cannot claim any unlimited validity. [V6, D24, p 328]
\end{quote}
More recently, Irwin Shapiro talked about the variable speed of light in his 1964 paper on what is now called the Shapiro delay.\cite{shapiro}
All these statements imply that light speed is not constant. It is only approximately so for regions within a nearly uniform gravitational potential.
These statements also suggest that the medium for light transmission is related to the dominant gravitational potential in the region of space under observation.
Thus, if we accept that gravity affects light speed, we are in essence accepting a form of the entrained ether, because all material bodies carry their gravitational potential along with them in their motion.

There is some evidence to support this in the experiment of Wolf and Petit that measured directional dependence of light speed using the satellites of the Global Positioning System (GPS).\cite{exp} For those who are unfamiliar with the GPS, this system use satellites about 20,000 km above the Earth’s surface, equipped with synchronized atomic clocks, and transmitters and receivers to broadcast signals carrying information of both times and positions.
It or systems like it used an Earth-Centered Inertial (ECI) frame to calculate the positions and times of moving transmitters or receivers to high accuracy.\cite{gps}
The ECI frame is an approximate inertial frame so does not partake in the Earth’s rotation. Earth's gravity dominates this frame within the altitude of the GPS satellites. It is where most experiments are conducted, including those of Wolf and Petit.
Their experiment gave an upper limit to light speed anisotropy of $\delta c/c < 5 \times 10^{-9}$, and thus showed that light speed is constant and independent of direction to a very high degree for receivers {\it at rest} in this ECI frame.\cite{exp} 

Those interested in modern entrained ether models may want to investigate the works of C.C. Su and C. K. Thornhill. \cite{su, thornhill, thornhill2, thornhill3, thornhill4, thornhill5, thornhill6}  They developed the entrained model theory using a classical Newtonian framework in two different ways. Su’s model even incorporated gravitational potential in its formulas/

Let us return to the c-equivalence principle. The same argument used to derive \eref{forward}, \eref{back}, and \eref{twowayspeed} can be applied to a transceiver moving relative to the ECI frame with velocity $\vect{u}$.
Here $c$ would play the role of $c_o$, and the two-way speed of light as measured by the moving transceiver, $c_m$, would be related to $c$ by:
\begin{equation} \label{speed2}
c_m = \frac{ c ( 1 -\beta^2)}{( 1 -\beta^2 \sin^2 \phi)^{1/2} }, \quad \beta = \frac{u}{c},
\end{equation}
with $\phi$ being the angle between $\vect{u}$ and the direction of the light path. 

Some researchers have proposed experiments to test the formulas corresponding to \eref{forward} and \eref{back} where $\phi = 0$ or $\phi = \pi$, with modern technologies and the global positioning system. \cite{hatch, wang} Among them is Ronald Hatch, a GPS expert who held various patents on GPS technologies as either inventor or co-inventor and who served numerous roles within the Institute of Navigation (ION).\cite{hatch}
These experiments would disprove the second hypothesis of STR--interpreted in its strong form that the speed of light is the same in all inertial frames. It would also disprove the c-equivalence hypothesis, so that the two-way kinematic speed of light, its wavefront, or group speed, would depend on the velocity of the moving detector with respect to the ether frame.
With the speed of light not being the same in every inertial frame, the composition of two Lorentz transformations would not yield a Lorentz transformation, as an elementary computation will reveal.
It would also invalidate some derivations of the Lorentz transformation.

Since the Earth also rotates, in actual earth-based experiments conducted over several days, its velocity $\vect{v}$ relative to the ether frame is not constant from hour to hour, and so $c$ would have minor daily variations.
Such daily speed of light variations could appear in interferometry experiments like the ones of Michelson and Morley and the ones of Kennedy and Thorndike. We will briefly touch upon these experiments in the next section.

\section{Michelson-Morley, Kennedy-Thorndike and Ives-Stilwell Type Experiments}

Three famous experiments are often said to provide evidence for the Lorentz transformation, mainly because we can use them to derive the LT. Robertson was the first person to show that their results imply the Lorentz transformation.\cite{robertson} These experiments are the Michelson-Morley (M-M), Kennedy-Thorndike (K-T), and Ives-Stilwell (I-S) experiments.  We will examine the assumption and conclusions drawn from these three experiments to see if they have any bearing on the c-equivalence principle. 

Albert Michelson was a firm believer in the ether theory and designed some experiments to detect the relative orbital motion of the Earth through the ether. To implement his plan, Michelson created a device now called an interferometer. This device splits the beam from a single light source with a half-silvered mirror and sends the two beams traveling at right angles to one another. After leaving the splitter, the beams traveled out to the ends of arms of nearly equal lengths, where small mirrors reflect them to the middle. These beams would recombine on the far side of the splitter to produce a pattern of constructive and destructive interference. Any slight change in the beams' travel time due to a change in speed of light would then be observed as a shift in the positions of the interference fringes. The actual light path in real experimental setup-ups is more complicated than what we described above. In 1887, Michelson and Edward Morley conducted their first ether wind detection experiment together at the Physics Department of Case School of Applied Science (today Case Western Reserve University) and got a result that was much smaller than expected, ``4 - 8 km/s’’ instead of ``30 km/s’’ and was declared to be zero. \cite{m-m}

Some researchers think that the shift is not zero when done at a high enough altitude, often quoting Dayton Miller's interferometry experiments as evidence. In his 1933 paper, Miller comprehensively summarized his interferometry work and included the data that supported his conclusions. Miller’s experimental setup made over 200,000 individual readings, over 12,000 turns of the interferometer undertaken at different months of the year and different altitudes, starting in 1902 at the Case School and ending in 1926 with his Mt. Wilson experiments. After having plotted the data against sidereal time, Miller remarked ``...a very striking consistency of their principal characteristics...azimuth and magnitude... as though they were related to a common cause... The observed effect is dependent upon sidereal time and is independent of diurnal and seasonal changes of temperature and other terrestrial causes, and...is a cosmical phenomenon.’’ \cite[p 231]{miller} 

After Miller’s death, his former student Shankland and several other people, in consultation with Einstein, reanalyzed some of Miller’s experimental data. They concluded that there must have been a thermal effect, but gave no evidence for it. The team admitted that it ``...did not embark on a statistically sound recomputation of the cosmic solution, but... [searched for]...local disturbances such as may be caused by mechanical effects or by nonuniform temperature distributions in the observational hut.’’ \cite[p 172]{shank}  With this admission, their paper has little scientific value. Fortunately, Case school archived Miller’s data, so a more rigorous and systematic reanalysis is possible.

Thomas Roberts did this analysis and made it available for all to consider. He concluded: 
\begin{quote}
	...[Miller] was unknowingly looking at statistically insignificant patterns in his systematic drift that mimicked the appearance of a real signal. An upper limit on ``absolute motion'' of 6 km/sec is derived from his raw data, fully consistent with similar experimental results ... The key point of this paper is the need for a comprehensive and quantitative error analysis. The concepts and techniques used in this analysis were not available in Miller’s day, but are now standard. These problems also apply to the famous measurements of Michelson and Morley, and to most if not all similar experiments ...\cite{roberts}
\end{quote}

More recently, Consoli, Matheson and Pluchino reanalyzed the Michelson-Morley, Miller, Joos, and several other classical ether-drift experiments.\cite{EDriftExp} They assumed the existence of a preferred frame called the ether, which they identified as the Cosmic Microwave Background frame. They also include in their model the refractive index of the gases used in these experiments. Consoli et al. concluded 
\begin{quote}
	... by introducing the refractive index $N$ of the gas, convective currents of the gas molecules would produce a small anisotropy, proportional to $(N - 1)(v/c)^2$, of the two-way velocity of light ... In the old times, experiments were performed with interferometers where light was propagating in gaseous media, air or helium at atmospheric pressure, where $N - 1$ is a very small number. In this regime, the theoretical fringe shifts expected ... are much smaller than the classical prediction $(v/c)^2$...
	These arguments make more and more plausible that a genuine physical phenomenon, much smaller than expected and characterized by stochastic variations, might have been erroneously interpreted as an instrumental artifact thus leading to the standard `null interpretation' of the experiments reported in all textbooks.\cite[p 49-50]{EDriftExp}
\end{quote}

In our estimate, these experimental results are inconclusive at best, but let us explore the logical ramification of a null result from the M-M type experiments. A null result only eliminates version (1A) of the ether theory, that the Earth moved through the ether without dragging it, and not that the ether is nonexistent, as many often claimed. To explain the null result, Lamor and Lorentz hypothesized that motion contracted the length of the arm parallel to the direction of the velocity by just the right amount. Thus, many claimed that the M-M null experimental results provide evidence for length contraction due to motion. This is an obvious example of circular reasoning, and the only way to break the circle is to have some independent evidence of length contraction. For emphasis, we repeat that experiments of the M-M type when done near the Earth's surface, only revealed that its speed relative to the ether is close to zero near the surface, but does not eliminate other theories. 

Even if we reject entrained ether models, we must explain why the ballistic theory of light is ruled out before we can claim that the null result of the M-M experiments is due to length contraction. Yet, if we accept light beams are streams of photons with wavelike behaviors emerging from large ensembles, we are adopting a form of the ballistic theory of light. In either of the above cases, there is no need for the length contraction hypothesis to explain the null results of M-M type experiments.

Let us now examine the Kennedy-Thorndike (K-T) type experiments. They are a variant of M-M experiments, with the distinguishing feature being an interferometer whose arms are of unequal lengths.\cite{k-t} Because of the unequal arms’ lengths, their experimental apparatus acted like an optical gyroscope and could detect the rotation of the Earth, as mentioned in their paper. The most important point to consider is the havefollowing. Without length contraction, K-T type experiments are also inconclusive at best, since these experiments assumed length contraction to prove so-call time dilation. In the words of Kennedy and Thorndike:
\begin{quote}
	The Michelson-Morley experiment indicates that a system moving with uniform velocity $v$ with respect to such a system has dimensions in the direction of motion contracted in the ratio $[ 1 - v^2/c^2 ]^{1/2}$ as compared to dimensions in the fixed system, while dimensions in the direction of and perpendicular to the motion are unchanged. This is in part assumption ... nevertheless it actually shows only that dimensions in the direction of and perpendicular to the motion are in the ratio mentioned; either of these dimensions might be any function of the velocity so long as that ratio is preserved. \cite{k-t}
\end{quote}

Both of these types of experiments give results smaller than expected near the Earth’s surface, so many researchers declared their results to be null. If we accept the entrained ether model, then Wolf and Petit experiment \cite{exp} implies that we must conduct these experiments in orbit above the altitude of the GPS satellites to get non-null results. In truth, null result experiments can only eliminate theories and do not support any theory. In this sense, they can be used to support all sorts of theories, as was pointed out by many people, including Herbert Ives. Specifically, he showed that any theory that has length contraction in the direction of motion in the ratio
\begin{equation}
[ 1 - v^2/c^2]^{(n+1)/2}: 1,
\end{equation}
and in the perpendicular direction of the motion by
\begin{equation}
[ 1 - v^2/c^2]^{n/2}: 1,
\end{equation}
would give a null result for the M-M experiment.
The null result of the K-T experiment is then explained if we alter the time scale at the origin in the ratio
\begin{equation}
[ 1 - v^2/c^2]^{(1-n)/2}: 1.
\end{equation}
Ives argued that the M-M and K-T experiments could not determine $n$, but his type of experiments could. \cite{ives37B}

To give some context to the Ives-Stilwell experiment, we need to describe his mindset. Herbert Ives was a firm believer in the Lorentz-Larmor relativity theory, an ether-based theory that uses different premises from those of Einstein to get the Lorentz transformation. He did not accept the constancy of light speed hypothesis and even thought that moving detectors would get different values depending upon their speed.\cite{ives37B} Let us turn to the last famous experiment. Unlike the previous two, the Ives-Stilwell type experiments give positive results. In our times, many people think experiments of the Ives-Stilwell type measure the traverse Doppler effect, often called time dilation. These experiments used moving light-emitting ion beams in various configurations and determined the shift in wavelength of the light. This shift seems to conform to a formula gotten by using the assumption of time dilation. In truth, experiments can only show that the data is consistent with a certain formula. If another theory can yield the same formula, experiments can not distinguish between them. Indeed, another way to get the same Doppler formula as the ones from STR exists. Schrodinger got the same equation in 1922 via a photon model and the laws of conservation of energy and linear momentum.\cite{quanta, dop4Photon} 

What conclusion can be obtained from the I-S experiments? We will quote someone who did a rigorous analysis, Wallace Kantor, to express our view. Not only did Kantor analyze the Ives-Stilwell experiments \cite{ives38, ives41}, he also evaluates the Otting experiment \cite{otting} and the Mandelberg-Witten experiment.\cite{man-wit} He concluded they were inconclusive at best and expressed the opinion `` It is very doubtful if the formidable technical difficulties inherent in these ion beam experiments can be overcome.’’ \cite{kantor}
We agree with Kantor, and add that if the result of Ives-Stilwell is accepted, it would actually falsify the c-equivalence principle. As Monti pointed out, Ives and Stilwell claimed that their experiment showed the following equations to hold for the wavelength:\cite[p 216]{ives38}
\begin{equation} \label{wav}
\lambda = \lambda_o ( 1 - v^2/c^2)^{1/2},
\end{equation}
and for the frequency, \cite[p 226]{ives38}
\begin{equation} \label{freq}
\nu = \nu_o ( 1 - v^2/c^2)^{1/2}.
\end{equation}
When multiplied together, we get the equation:
\begin{equation}
c =  c_o (1 - v^2/c^2)^{1/2},
\end{equation}
with $c= \lambda \nu$, $c_o = \lambda_o \nu_o$. Thus, if $c=c_o$ and they are finite, we must have $v=0$. This contradicts the motion of the ions in their experiment. If more recent Ives-Stilwell type experiments claimed to prove \eref{wav} and \eref{freq}, then they are showing that the c-equivalent principle is false.

\section{Proposed Experiments}

To falsify the c-equivalence principle, someone would need to conduct an experiment that shows $c \neq c_o$ beyond all doubts. This requires measuring $c_0$ and $c$ independently, then comparing them. There are some experimental indications that $c \neq c_o$ from the speed of radar signals that were bounced off of Venus,\cite{sat} Mars,\cite{sat1} interplanetary spacecrafts,\cite{sat2} and other interplanetary objects.\cite{sat3}
When Bryan Wallace analyzed the data from Venus, the data suggests that the classical velocity addition law holds for light, so that $c$ is not constant.\cite{wallace}
An independent scientist from Russia, Tolchel’nikova confirmed Wallace’s observations and presented her finding at an international conference.\cite{tol} With more recent radar experiments on spacecrafts using an observation window of several months, the measured radar echo time of the spacecrafts and the theoretical formula agreed with each other to within $1 \mu s$ or better.\cite{sat1, sat2} The theoretical formula came from a classical medium propagation model with a heliocentric inertial frame as the standard of rest for the radar signal. Models using other frames, such as the Earth-centered inertial or Earth-centered Earth-fixed frame, did not fit the data. The important implication for us is that the Earth’s nonzero rotational and orbital velocities with respect to this frame have to be added or subtracted away to get this agreement. So the first step to determine the truth of the $c$-equivalence principle should be a reanalysis of these interplanetary radar echo data. If these reanalyses are inconclusive, then more controlled experiments would be required. We considered these next. 

Currently, there are several methods to determine light speed, such as using cavity resonance and interferometry. The most relevant to us is measuring the electromagnetic constants $\epsilon_o$ and $\mu_o$, and time-of-flight methods.
The vacuum permittivity $\epsilon_o$ is determined from measuring the capacitance and dimensions of a capacitor, while the value of the vacuum permeability is fixed at exactly $\mu_o = 4 \pi \times 10^{-7}$ H m$^{-1}$ through the definition of the ampere.
Rosa and Dorsey, the last recorded researchers to use EM measurements, used it in 1907 to get their result. \cite{rosa}
To determine if (\ref{twowayspeed}) hold, experimentalists would need to update the measure of $\epsilon_o$ with modern technology and perform it again along with modernized time-of-flight measurements.

The last recorded time-of-flight measurement was the experiment designed by Albert Michelson, of the famous M-M experiment, and carried out by Pease and Pearson between September 1929 and March 1933 in California.
In this experiment, a system of mirrors folded the path length of 12.8 to 16 km to fit within a 1.6 km long pipe, evacuated to pressures between 66 and 734 pascals.
During its lifetime, it made 2885.5 determinations of light speed, giving the simple mean of $c=$ 299,774 km/sec with an average deviation of 11 km/sec.\cite{mpp}
Since they did this experiment during different times of the day and year, variations due to the direction and speed of the Earth relative to the ether may exist within their experimental data. A reanalysis of their experimental data is worth doing to see if its variations can be accounted for.
Methods that use interferometry to measure wavelength and independent ways to measure frequency yielding $c$ as their product could replace the time-of-flight method, as long as they do not use standing waves.

Modern kinematic measures of $c$ can be done using the satellites of a global navigation satellite system like GPS, or a modernized version of the Michelson, Pease, and Pearson experiment.
If a modernized Michelson, Pease, and Pearson setup is used, we suggest having at least two identical devices far from urban centers and seismically active areas. We should also enclose them in climate control housing at a constant temperature. This would minimize error sources like temperature fluctuations and ground vibrations.
One such experimental setup should be at sea level and one at a higher altitude, perhaps at the top of a mountain.
The experimental apparatus should automatically send the raw daily data to independent labs for analysis. Experimenters would need to consider the CGPM (1983) definition of the meter in terms of the speed of light as pointed out by Mare and collaborators.\cite{shadow}

\section{Closing remarks}
What are the consequences of falsifying the c-equivalence principle? As we mentioned previously, showing that $c \neq c_o$ would invalidate the second postulate of the Special Theory of Relativity in its strong form interpretation that light speed has the same value in all inertial frames.
If someone shows that \eref{twowayspeed} hold, then the ether exists! And we found a way to determine speed relative to the ether, although it may need to be done far enough away from the dominant gravity field. This would also imply that the forward time of travel is greater than the return time $\Delta t_{f} > \Delta t_{r}$, so that the kinematic speed of light is no longer isotropic with respect to moving detectors.
Also, the transformations to preserve covariancy from one frame to another for the various electromagnetic fields will need to be changed. Since the Lorentz transformations of different frames depend on different $c$ values, the composition of two Lorentz transformations is rendered meaningless.
It would also invalidate any conclusion based upon the assumption that the two-way velocity of light is the same in all directions and all inertial systems, such as the Selleri transform. \cite[p 326]{selleri}
What is this ether? Many think it is the frame of the fixed stars or the frame wherein the cosmic microwave background radiation is uniform in all directions.\cite{ether} Others think it is the quantum vacuum.\\

\textbf{Acknowledgment} We would like to express our appreciation to Steffen Kuhn for his comments and suggestions to improve the paper. \\


\textbf{References} \\

\end{document}